# Radial Velocity Retrieval for Multichannel SAR

# Moving Targets with Time-Space Doppler De-ambiguity

Jia Xu[1*], Zu-Zhen Huang[1*], Zhi-Rui Wang[2], Li Xiao[3], Xiang-Gen Xia[1,3] and Teng Long[1]

(1: School of Information and Electronics, Beijing Institute of Technology, Beijing 100081, China)

(2: Department of Electronics Engineering, Tsinghua University, Beijing 100084, China)

(3: Department of Electrical and Computer Engineering, University of Delaware, Newark, DE 19716, USA)

**Corresponding authors:**

Jia Xu/Zu-Zhen Huang

School of Information and Electronics,

Beijing Institute of Technology, Beijing, 100081, P. R. China

**Tel & Fax:** +86-010-68915083

**E-mail:** xujia@bit.edu.cn/hzzhit@126.com

# Radial Velocity Retrieval for Multichannel SAR Moving Targets with Time-Space Doppler De-ambiguity

Jia Xu[1*], Zu-Zhen Huang[1*], Zhi-Rui Wang[2], Li Xiao[3], Xiang-Gen Xia[1, 3], and Teng Long[1]

(1: School of Information and Electronics, Beijing Institute of Technology, Beijing 100081, China)

(2: Department of Electronics Engineering, Tsinghua University, Beijing 100084, China)

(3: Department of Electrical and Computer Engineering, University of Delaware, Newark, DE 19716, USA)

xujia@bit.edu.cn/hzzhit@126.com

*Abstract*- In this paper, with respect to multichannel synthetic aperture radar (SAR), we first formulate the problems of Doppler ambiguities on the radial velocity (RV) estimation of a ground moving target in range-compressed domain, range-Doppler domain and image domain, respectively. It is revealed that in these problems, cascaded time-space Doppler ambiguity (CTSDA) may arise, i.e., time domain Doppler ambiguity (TDDA) in each channel arises first and then spatial domain Doppler ambiguity (SDDA) among multi-channels arises second. Accordingly, the multichannel SAR systems with different parameters are investigated in three cases with different Doppler ambiguity properties. Then, a multi-frequency SAR is proposed for RV estimation by solving the ambiguity problem based on Chinese remainder theorem (CRT). In the first two cases, the ambiguity problem can be solved by the existing closed-form robust CRT. In the third case, it is found that the problem is different from the conventional CRT problem and we call it a double remaindering problem in this paper. We then propose a sufficient condition under which the double remaindering problem, i.e., the CTSDA, can also be solved by the closed-form robust CRT. When the sufficient condition is not satisfied, a searching based method is proposed. Finally, some results of numerical experiments are provided to demonstrate the effectiveness of the proposed methods.

***Index Terms***- Multichannel synthetic aperture radar (SAR), ground moving target indication (GMTI), Chinese remainder theorem (CRT), double remaindering, radial velocity, time domain Doppler ambiguity, spatial domain Doppler ambiguity, cascaded time-space Doppler ambiguity.

## I. Introduction

It is known that ground moving target indication (GMTI) of synthetic aperture radar (SAR) has wide applications in both civilian and military fields [1-6]. For an uncooperative moving target, not only detection but also estimation, imaging, location and recognition should be accomplished for an advanced SAR-GMTI system. For the sake of these goals, the radial velocity (RV) of a moving target is an important motion parameter to be determined. In order to better suppress background clutter and estimate target motion parameters, a number of multichannel SAR systems have been proposed in the literature, such as displaced phase center antenna SAR (DPCA-SAR) [7-9], along track interferometry SAR (ATI-SAR) [10-13], space-time adaptive processing SAR (STAP-SAR) [14-17] and velocity SAR (VSAR) [18-22]. By combining the image formation in multi-channels, each multichannel SAR aforementioned can be flexibly implemented in range-compressed domain, range-Doppler (RD) domain or focused image domain. For example, in [22], the VSAR-based azimuth shift correction using an actual airborne system is first demonstrated with a real experiment.

Unfortunately, no matter which domain is used for multichannel processing in a SAR-GMTI system, there exist two kinds of Doppler ambiguities that can seriously degrade the estimation performance of a moving target's RV. First, as the azimuth signal in each channel is pulse-by-pulse sampled by pulse repetition frequency (PRF), Doppler ambiguity may arise in the slow time domain for a fast moving target, which is called time domain Doppler ambiguity (TDDA) in this paper. Second, the RV estimation based on the interferometric phase among multi-channels is influenced by the phase modulo folding. In other words, when the interferometric phase of a fast moving target is outside the interval $(-\pi, \pi]$, target's RV will be folded so that it cannot be uniquely determined. We call it spatial domain Doppler ambiguity (SDDA) in this paper, which is closely related to the "azimuth location ambiguity" in [23-25].

Over the past decades, many methods have been deliberated to deal with the TDDA problem. An intuitive method is to increase the PRF [26], but it will reduce the unambiguous SAR imaging swath and increase the computational complexity. In [27], a nonuniform PRF system was proposed to solve TDDA based on Chinese remainder theorem (CRT), but it requires a non-conventional pulse scheduling and increases the complexity in image formation. Besides, some other methods were proposed based on the envelope responses of moving targets [28-31], which can accomplish de-ambiguity of TDDA, but largely depend on the accurate measurements of a target's position and amplitude. Unfortunately, without the clutter cancellation processing among multi-channels, the direct

TDDA de-ambiguity and RV estimation methods may be affected by the background clutter. Therefore, as mentioned above, multichannel SAR-GMTI systems, like DPCA-SAR, ATI-SAR, STAP-SAR and VSAR, have been widely used for background clutter suppression. However, they will suffer from the SDDA problem. Then, multi-frequency SAR (MF-SAR) [32], nonuniform linear antenna array SAR (NULA-SAR) [23], dual-speed SAR (DS-SAR) [24] and bistatic linear antenna array SAR [25] have been proposed based on CRT to solve the SDDA problem. Nevertheless, the real RV related to moving target motion cannot be retrieved only from SDDA de-ambiguity. Thus, a frequency diversity based ambiguity resolver has been discussed in [21], where different wavelengths with small differences can be used by range multi-look processing to estimate the unambiguous RV.

In this paper, based on the range-Doppler imaging of a static scene, a moving target's interferometric phase is first derived in range-compressed domain, RD domain and image domain for a general multichannel SAR system. It is found that a multichannel SAR system with different parameters can be divided into the following three cases. For a Case I system, the time sampling frequency is smaller than the space sampling frequency, and only the TDDA will arise. For a Case II system, the time sampling frequency is an integer multiple of the space sampling frequency, and the cascaded time-space Doppler ambiguity (CTSDA) will arise, that is, the ambiguous Doppler frequency of a fast moving target after TDDA will be measured again by spatial sampling among multi-channels. Fortunately, the CTSDA for Case II systems can be thought of as the SDDA. For a Case III system, the time sampling frequency is larger than the space sampling frequency but not an integer multiple of that, and the CTSDA will also arise. Then, an MF-SAR [32] is proposed for RV estimation by solving the ambiguity problem. Both ambiguity problems in Cases I and II systems can be simply solved by the closed-form robust CRT in [33, 34]. For Case III systems, the CTSDA problem is different from the conventional CRT problem. In a Case III system, two levels of modulo operations are involved, where an integer is first taken a modulo with a positive integer $M$ and then its remainder is taken another modulo with a positive integer $N$ with $N<M$. We call this problem a double remaindering problem. For the double remaindering problem, we first propose a sufficient condition for the uniqueness of a solution. With the proposed sufficient condition, the double remaindering problem can be degenerated to the conventional CRT problem and solved by the closed-form robust CRT. When the sufficient condition is not satisfied, a searching based reconstruction method is proposed similar to the existing robust CRT in [35, 36]. Based on numerical experiments and performance analysis, it is validated that the proposed MF-SAR can well accomplish the CTSDA de-ambiguity and obtain the unambiguous RV via the multiple ambiguous radial velocities in space domain with respect to multiple frequencies.

The remainder of this paper is organized as follows. In Section II, TDDA and SDDA in a multichannel SAR system are derived in range-compressed domain, RD domain and image domain, respectively. In Section III, based on the relationship between TDDA and SDDA, the SAR systems are divided into three cases. In Section IV, the MF-SAR is proposed to obtain the target's real RV, and two reconstruction algorithms are propounded for the double remaindering problem in a Case III system. In Section V, simulation results are provided to demonstrate the effectiveness of the proposed methods. Finally, the conclusions are drawn in Section VI.

## II. Doppler Ambiguities of SAR Moving Targets in Time and Space Domains

The geometry of an along-track multichannel SAR is shown in Fig. 1, where the $x$-axis is the along-track direction and the $y$-axis is the cross-track direction. The multichannel SAR is equipped with a linear antenna array of $M$ receiving antennas with a uniform spacing $d$, in which the $0^{th}$ antenna serves as both the transmitter and the receiver, and the other antennas only serve as the receivers. The radar platform flies along the azimuth direction at an altitude $h$ with a constant forward velocity $v_a$. The sampling time $t$ along the $x$-axis is the slow time while that along the $y$-axis, i.e., $\tau$, is the fast time. When $t=0$, the $0^{th}$ antenna is located at $(0,0,h)$, and a moving target $P$ is located at $(0,y_0,0)$. During the radar illumination, the target $P$ is assumed to move with constant cross-range velocity $v_x$ and range velocity $v_y$.

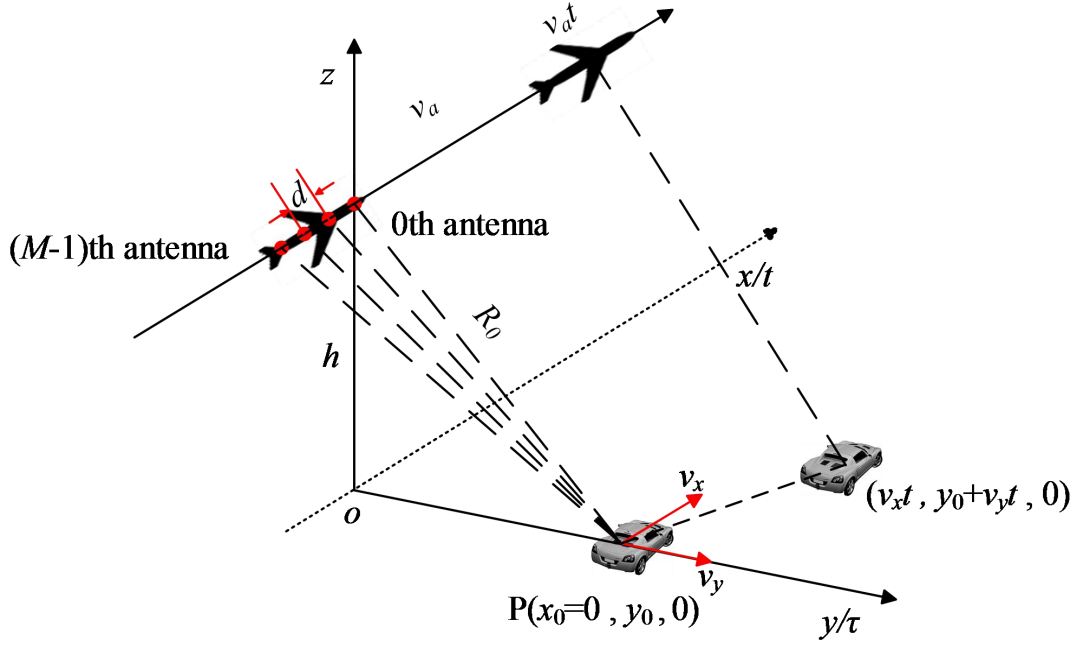

Fig. 1 Along-track multichannel SAR geometry.

Suppose the SAR transmits a linear frequency-modulated signal, i.e.,

$$s_o(\tau) = A \exp\left[ j2\pi\left( f_c\tau + \frac{\gamma}{2}\tau^2 \right) \right] \mathrm{rect}(\tau/T),$$  (1)

where $\mathrm{rect}(\cdot)$ is a rectangular function, and $A, f_c$, $\gamma = B/T$, $B$ and $T$ denote the amplitude, the carrier frequency, the pulse chirp rate, the pulse bandwidth and the pulse duration of the transmitting signal, respectively. Define $R_{0m}(t) = R_0(t) + R_m(t)$, where $R_{0m}(t)$ is the two-way range between target $P$ and the radar, and $R_m(t)$ is the instantaneous range from target $P$ to the $m^{\mathrm{th}}$ antenna:

$$R_m(t) = \sqrt{\left[ (v_a - v_x)t - md \right]^2 + (y_0 + v_y t)^2 + h^2}, \, m = 0,1,\cdots,M-1.$$  (2)

After demodulation, the base-band echo of target $P$ at the $m^{\mathrm{th}}$ antenna can be represented as

$$s(\tau,t,m) = \sigma\,\mathrm{rect}\left( \frac{\tau - R_{0m}(t)/c}{T} \right) \exp\left[ j\pi\gamma\left( \tau - \frac{R_{0m}(t)}{c} \right)^2 \right] \mathrm{rect}\left( \frac{t}{T_s} \right) \exp\left[ -j2\pi\frac{R_{0m}(t)}{\lambda} \right],$$  (3)

where $T_s$ is the target illumination time, $c$ is the speed of light, $\lambda = c/f_c$ is the wavelength, and $\sigma$ is the echo's constant amplitude.

After the range compression of (3) via matched filtering on wideband SAR echoes, the received signal can be well approximated as

$$s(\tau,t,m) = \sigma\sqrt{TB}\,\text{sinc}\left[B\left(\tau - \frac{R_{0m}(t)}{c}\right)\right]\text{rect}\left(\frac{t}{T_s}\right)\exp\left[-j\frac{2\pi R_{0m}(t)}{\lambda}\right]. \qquad (4)$$

For convenience, $R_{0m}(t)$ is approximated by its Taylor polynomial of degree 2:

$$R_{0m}(t) = R_0(t) + R_m(t) \approx 2R_0 - \lambda f_{D,m}t - \frac{\lambda}{2}f_{rT}t^2 + \Phi\,, \qquad (5)$$

where $R_0 = R_0(t)|_{t=0} = \sqrt{y_0^2 + h^2}$, $f_{rT} = -2\left[(v_a - v_x)^2 + v_y^2\right]/(\lambda R_0)$, $\Phi = m^2 d^2/(2R_0)$, and

$$f_{D,m} = -\frac{2y_0 v_y - (v_a - v_x)md}{\lambda R_0} = mf_0 + f_d\,, \qquad (6)$$

$$f_d = -\frac{2y_0 v_y}{\lambda R_0} = -\frac{2v_r}{\lambda}\,, \qquad (7)$$

where $f_0 = (v_a - v_x)d/(\lambda R_0)$ is the Doppler introduced by the antenna azimuth location and is usually small, $f_d$ is the Doppler caused by the target's radial motion, and $v_r = v_y y_0/R_0$ is the RV.

### A. Doppler Ambiguity in Range-Compressed Domain

Because the sampling frequency along slow time is PRF $f_P$, the measured Doppler frequency of a moving target is limited in $(-f_P/2, f_P/2]$. When the Doppler frequency of a fast target is outside this interval, the TDDA arises. From (6) the target's Doppler frequency $f_{D,m}$ in the $m$th channel can be expressed as

$$f_{D,m} = \hat{f}_{D,m} + N_T f_P = mf_0 + \hat{f}_d + N_T f_P \approx \hat{f}_d + N_T f_P\,, \qquad (8)$$

where $N_T$ is the folding integer of the TDDA, $\hat{f}_d = -2v_{r,\text{time}}/\lambda \in (-f_P/2, f_P/2]$ is the measured ambiguous Doppler frequency in $0$th channel with respect to the measured ambiguous RV $v_{r,\text{time}} \in [-V_T/2, V_T/2)$, and $V_T$ is the corresponding TDDA modulus, i.e., blind speed in time domain,

$$V_T = \lambda f_P/2\,. \qquad (9)$$

Thus, from (7), (8) and (9) the real RV of a moving target can be represented as

$$v_r = v_{r,\text{time}} + N_T V_T\,. \qquad (10)$$

Substituting (10) into (4), the signal after range compression can be expressed as

$$s(\tau,t,m) = \sigma\sqrt{TB}\,\mathrm{sinc}\!\left[B\!\left(\tau - \frac{R_{0m}(t)}{c}\right)\right]\mathrm{rect}\!\left(\frac{t}{T_s}\right)$$
$$\cdot\exp\!\left\{-j\frac{2\pi}{\lambda}\!\left(2R_0 + \frac{\left[2y_0 v_{y,\mathrm{time}} - (v_a - v_x)md\right]}{R_0}t + \frac{\left[(v_a - v_x)^2 + v_y^2\right]}{R_0}t^2 + \frac{m^2 d^2}{2R_0}\right)\right\}, \quad (11)$$

where $v_{y,\mathrm{time}} = v_{r,\mathrm{time}} R_0 / y_0$. According to (11), the phase of the signal after range compression is related to the folded RV $v_{r,\mathrm{time}}$ instead of the real RV $v_r$ due to TDDA, while the time-varied envelope $\mathrm{sinc}\!\left[B(\tau - R_{0m}(t)/c)\right]$ is still related to the real RV $v_r$ from (5).

After slow-time co-registration and phase compensation, the interferometric term between the $0^{\mathrm{th}}$ channel and the $m^{\mathrm{th}}$ channel in range-compressed domain can be given as

$$\Delta P_{\mathrm{rc}}(m) = s^*(\tau,t,0)\cdot\left(s\!\left(\tau,t+\frac{md}{2v_a},m\right)\cdot P_{\mathrm{rc},c}\right)$$
$$= \exp\!\left\{-j\frac{2\pi}{\lambda}\!\left[\!\left(\frac{(v_a - v_x)^2 + v_y^2}{R_0}\cdot\frac{md}{v_a} - \frac{(v_a - v_x)md}{R_0}\right)t\right.\right.$$
$$\left.\left. + \frac{y_0 v_{y,\mathrm{time}}md}{R_0 v_a} + \frac{(v_a - v_x)^2 + v_y^2}{R_0}\cdot\frac{m^2 d^2}{4v_a^2} - \frac{(v_a - v_x)m^2 d^2}{2R_0 v_a} + \frac{3m^2 d^2}{4R_0}\right]\right\} \quad , \quad (12)$$

where $(\cdot)^*$ is the complex conjugate operator, and $P_{\mathrm{rc},c} = \exp\!\left[j\pi m^2 d^2 / (2\lambda R_0)\right]$ is the phase compensation function. Since $v_x \ll v_a$, $v_y \ll v_a$, and $R_0$ is large enough, (12) can be well approximated as

$$\Delta P_{\mathrm{rc}}(m) \approx \exp\!\left(-j\frac{2\pi y_0 v_{y,\mathrm{time}}md}{\lambda R_0 v_a}\right) = \exp\!\left(-j\frac{2\pi v_{r,\mathrm{time}}md}{\lambda v_a}\right). \quad (13)$$

Let $\Delta_s = d/(2v_a)$ denote the sampling interval in space domain. Then, from $\hat{f}_d = -2v_{r,\mathrm{time}}/\lambda$ we can express (13) as $\Delta P_{\mathrm{rc}}(m) \approx \exp\!\left(j2\pi\hat{f}_d m\Delta_s\right)$. When $v_{r,\mathrm{time}}$ is so large that $\hat{f}_d$ is outside the interval $(-F_s/2, F_s/2]$, the SDDA arises, i.e.,

$$\hat{f}_d = f_{\mathrm{space}} + N_S F_S, \quad (14)$$

where $F_s = 1/\Delta_s$ is the space sampling frequency, $N_s$ is the folding integer of the SDDA,

$$f_{\mathrm{space}} = -2v_{r,\mathrm{space}}/\lambda \in (-F_s/2, F_s/2] \quad (15)$$

is the measured ambiguous Doppler frequency in space domain with respect to the measured ambiguous RV $v_{r,\text{space}} \in \left[-V_S/2, V_S/2\right)$, and $V_S$ is the corresponding SDDA modulus, i.e., blind speed in space domain,

$$V_S = \lambda F_s/2 = \lambda v_a/d .\qquad (16)$$

Then, the interferometric term in range-compressed domain can be expressed as

$$\Delta P_{\text{rc}}(m) = \exp\left[j2\pi\left(f_{\text{space}} + N_S F_S\right)m\Delta_s\right] = \exp\left(j2\pi f_{\text{space}} m\Delta_s\right) = \exp\left(-j\frac{2\pi m d v_{r,\text{space}}}{\lambda v_a}\right).\qquad (17)$$

This means that $v_{r,\text{space}}$ is the obtained target's RV in range-compressed domain based on the interferometric phase. From (14) and the Doppler definition in (8), $v_{r,\text{space}}$ can be expressed as

$$v_{r,\text{time}} = v_{r,\text{space}} + N_S V_S .\qquad (18)$$

Although (10) and (18) show the CTSDA problem in range-compressed domain, one might wonder what will happen in other domains since SAR-GMTI and RV estimation can be realized in different domains, e.g., range-Doppler (RD) domain [29] and image domain [15, 19], in the process of SAR image formation.

Next we will elaborate what happens in the RD and image domains, respectively.

## B. Doppler Ambiguity in RD Domain

After the Fourier transform (FT) of (11) along $t$, we get

$$s_{RD}(\tau, f_t, m) = \frac{\sigma\sqrt{TB}}{\sqrt{|f_{rT}|}}\text{sinc}\left\{B\left[\tau - \frac{2}{c}\left(R_0 - \frac{\lambda}{4}\left(\frac{f_t^2 - \hat{f}_{D,m}^2 + 4N_T f_P\left(f_t - \hat{f}_{D,m}\right)}{f_{rT}}\right)\right)\right]\right\}$$
$$\cdot \text{rect}\left(\frac{f_t - \hat{f}_{D,m}}{B_d}\right)\exp\left[-j\pi\frac{\left(f_t - \hat{f}_{D,m}\right)^2}{f_{rT}}\right]\exp\left(j\left(\frac{2\pi\Phi}{\lambda} + \varphi_0\right)\right),\qquad (19)$$

where $f_t$ is the Doppler frequency, $B_d = T_s f_{rT}$, and $\varphi_0 = 4\pi R_0/\lambda - \pi/4$ is a constant phase. For the conventional SAR imaging, the range-migrations of both static and moving targets are corrected according to that of a static target. The target's signal after range cell migration correction (RCMC) in RD domain is given as

$$s_{RD}\left(\tau,f_t,m\right)=\frac{\sigma\sqrt{TB}}{\sqrt{\left|f_{rT}\right|}}\mathrm{sinc}\left\{B\left[\tau-\frac{2}{c}\left(R_0-k_r\left(f_t-\hat{f}_{D,m}\right)-r_a\right)\right]\right\}\mathrm{rect}\left(\frac{f_t-\hat{f}_{D,m}}{B_d}\right)$$
$$\cdot\exp\left[-j\pi\frac{\left(f_t-\hat{f}_{D,m}\right)^2}{f_{rT}}\right]\exp\left(j\left(\frac{2\pi\Phi}{\lambda}+\varphi_0\right)\right) \quad , \quad (20)$$

where $k_r=\lambda N_T f_P/\left(2f_{rT}\right)$, and $r_a=-\lambda f_{D,m}^2/\left(4f_{rT}\right)$. From (20) it can be observed that the phase in RD domain is also related to the folded RV after TDDA instead of the real RV, while the amplitude response approximately follows a straight line, whose slope is primarily determined by the folding integer of the TDDA $N_T$.

Then, the interferometric term between the $0^{\mathrm{th}}$ channel and the $m^{\mathrm{th}}$ channel in RD domain can be expressed as

$$\Delta P_{RD}\left(m\right)=s_{RD}^*\left(\tau,f_t,0\right)\cdot s_{RD}\left(\tau,f_t,m\right)$$
$$=\exp\left\{j\pi\left[\frac{m^2d^2\left(v_a-v_x\right)^2}{2\lambda R_0\left(\left(v_a-v_x\right)^2+v_y^2\right)}-\frac{md\left(v_a-v_x\right)}{\left(v_a-v_x\right)^2+v_y^2}f_t-\frac{2mdy_0v_{y,\mathrm{time}}\left(v_a-v_x\right)}{\lambda R_0\left(\left(v_a-v_x\right)^2+v_y^2\right)}\right]\right\}. \quad (21)$$

Since $v_x<<v_a$, $v_y<<v_a$, and $R_0$ is large enough, the first exponential component in (21) can be approximated as zero, and the second exponential component can be approximated as $-f_t md/v_a$ that corresponds to the time registration in range-compressed domain and is needed to be compensated. So the interferometric term in RD domain can be approximated as

$$\Delta P_{RD}\left(m\right)=\exp\left(-j\frac{2\pi mdv_{r,\mathrm{time}}}{\lambda v_a}\right)=\exp\left(j2\pi f_{\mathrm{space}}m\Delta_s\right)=\exp\left(-j\frac{2\pi mdv_{r,\mathrm{space}}}{\lambda v_a}\right), \quad (22)$$

which is identical to the interferometric term in the range-compressed domain. That is, the obtained target's RV in RD domain based on the interferometric phase in (22) is also $v_{r,\mathrm{space}}$, and the CTSDA problem in RD domain is the same as that in (10) and (18).

## C. Doppler Ambiguity in SAR Image Domain

After RCMC, in order to fulfill image co-registration and miniature error elimination for (20), a compensation function in RD domain is given as

$$P_{RD,c}\left(\tau,f_t,m\right)=\exp\left(-j2\pi mf_t t_d\right)\exp\left[j\pi\left(m^2f_0t_d-\frac{2\Phi}{\lambda}\right)\right], \quad (23)$$

where $t_d=d/2v_a$. After (20) is multiplied by (23), a Doppler compensation function $\exp\left(-j\pi f_t^2/f_r\right)$ with respect to a static target is utilized to implement the azimuth focusing, where $f_r=-2v_a^2/\left(\lambda R_0\right)$ is the Doppler rate of the static target. Due to the uncompensated motion of a moving target, its ultimate image can be classified into three types [37] in accordance with different radial and cross-range velocities. With $\delta=4\sqrt{v_a^2-v_y^2}/\left(\lambda R_0\right)$, $v_0=v_a-\sqrt{v_a^2-v_y^2}$, $\rho_1=T_s^2$ and $\rho_2=\left[c/\left(\lambda BN_Tf_P\right)\right]^2$, the above three types can be given as follows [37].

**Type I**: When $N_T=0$ and $\left|v_x-v_0\right|\le 1/\left(\delta\rho_1\right)$, the time-bandwidth product (TBP) of (20) after Doppler compensation will approximately be 1. By performing inverse FT along $f_t$, the signal response in the image domain can be approximated as

$$
\begin{aligned}
s_{\text{image,I}}\left(\tau,t,m\right)\approx\sigma_1\,\text{sinc}&\left\{B\left[\tau-\frac{2}{c}\left(R_0-r_a\right)\right]\right\}\cdot\text{sinc}\left[B_d\left(t+\frac{\hat{f}_d}{f_r}\right)\right] \\
&\cdot\exp\left(j2\pi\hat{f}_{D,m}t\right)\exp\left(j2\pi m\hat{f}_dt_d\right)
\end{aligned}
\tag{24}
$$

where $\sigma_1$ is a complex-valued constant.

**Type II**: When $N_T=0$ and $\left|v_x-v_0\right|>1/\left(\delta\rho_1\right)$, or $N_T\ne0$ and $\left|v_x-v_0\right|>1/\left(\delta\rho_2\right)$, the TBP of (20) after Doppler compensation will be far larger than 1. By performing inverse FT along $f_t$ via stationary phase principle (SPP), the signal response in the image domain can be approximated as

$$
\begin{aligned}
s_{\text{image,II}}\left(\tau,t,m\right)\approx\sigma_2\,\text{rect}&\left[\frac{f_rt+\hat{f}_d}{\left(f_r-f_{rT}\right)T_s}\right]\text{sinc}\left\{B\left[\tau-\frac{2}{c}\left(R_0-r_a\right)+\frac{2}{c}k_{\,f}r_{rT}\frac{f_rt+\hat{f}_d}{f_r-f_{rT}}\right]\right\} \\
&\cdot\exp\left[j\pi\frac{f_rf_{rT}}{f_r-f_{rT}}\left(t+\frac{\hat{f}_d}{f_r}\right)^2\right]\exp\left(j2\pi\hat{f}_{D,m}t\right)\exp\left(j2\pi m\hat{f}_dt_d\right)
\end{aligned}
\tag{25}
$$

where $\sigma_2$ is a complex-valued constant.

**Type III**: When $N_T\ne0$ and $\left|v_x-v_0\right|\le1/\left(\delta\rho_2\right)$, the TBP of (20) after Doppler compensation will approximately be 1. By performing inverse FT along $f_t$, the signal response in the image domain can be approximated as

$$s_{\text{image,III}}\left(\tau,t,m\right) \approx \sigma_3 \, \text{rect}\left[\frac{c\tau-2\left(R_0-r_a\right)}{2k_r B_d}\right]\text{sinc}\left[\frac{\pi c}{2k_r B}\left(t+\frac{\hat{f}_d}{f_r}+f_{dc}\left(\tau\right)\right)\right],$$
$$\cdot \exp\left(j2\pi \hat{f}_{D,m}t\right)\exp\left(j2\pi m\hat{f}_d t_d\right) \tag{26}$$

where $\sigma_3$ is a complex-valued constant and

$$f_{dc}\left(\tau\right) = \left[\frac{c\left(f_r-f_{rT}\right)}{2f_r f_{rT}k_r}\right]\cdot\left[\tau-\frac{2\left(R_0-r_a\right)+k_r\hat{f}_d}{c}\right]. \tag{27}$$

The derivations of (24)-(27) can be found in [37]. According to (24), (25) and (26), a moving target located at $t = 0$ is finally imaged at $t = -\hat{f}_d/f_r$ in the SAR image. In other words, the azimuth shift of a moving target in the image domain is

$$\Delta x = -\frac{\hat{f}_d}{f_r}v_a = -\frac{v_{r,\text{time}}}{v_a}R_0. \tag{28}$$

Notably, the azimuth shift is determined by the folded RV $v_{r,\text{time}}$ after TDDA rather than the real RV, which tells us that the moving target location in image domain can be obtained by only solving the SDDA. Therefore, SDDA is equivalent to "azimuth location ambiguity" [23-25]. Because $v_{r,\text{time}}$ is limited in $[-V_T/2, V_T/2)$, from (28) the maximum possible azimuth shift of a moving target is

$$\Delta x_{\max} = \frac{V_T}{2v_a}R_0 = \frac{\lambda f_p}{4v_a}R_0. \tag{29}$$

Furthermore, although moving targets are divided into three types [37], their interferometric terms between the $0^{\text{th}}$ channel and the $m^{\text{th}}$ channel are identical regardless of the target types, i.e.,

$$\Delta P_{\text{image}}\left(m\right) = \exp\left[j2\pi\left(\hat{f}_{D,m}-\hat{f}_{D,0}\right)t\right]\exp\left(j2\pi m\hat{f}_d t_d\right)$$
$$= \exp\left[-j2\pi\frac{\left(v_a-v_x\right)md}{\lambda R_0}t\right]\exp\left(-j\frac{2\pi mdv_{r,\text{time}}}{\lambda v_a}\right)\cdot \tag{30}$$

Since $v_x \ll v_a$ and $R_0$ is large enough, the first exponential component in (30) is approximated as zero. Therefore, the interferometric term can be approximated as

$$\Delta P_{\text{image}}\left(m\right) = \exp\left(-j\frac{2\pi mdv_{r,\text{time}}}{\lambda v_a}\right) = \exp\left(j2\pi f_{\text{space}}m\Delta_s\right) = \exp\left(-j\frac{2\pi mdv_{r,\text{space}}}{\lambda v_a}\right), \tag{31}$$

which is also identical to the interferometric term in range-compressed and RD domains. This means that $v_{r,\text{space}}$ is also the obtained target's RV in image domain based on the interferometric phase and the CTSDA problem in image domain is the same as that in (10) and (18).

In accordance with (11), (17), (22) and (31), it is revealed that a moving target may suffer from TDDA once the echoes are received. Then the ambiguous velocity in time domain may suffer from SDDA in space domain. That is, TDDA arises first and then SDDA arises subsequently for a moving target in different domains of SAR image formation.

### III. SAR System Classification From Different Doppler Ambiguities

From (11), (17), (22) and (31), one can see that the target's Doppler ambiguity is related to the real RV and system parameters, and can be divided into the following three cases.

*1) Case I:*

The condition of this case is $V_T < V_S$. From (9) and (16), this condition is equivalent to

$$d < 2v_a/f_P . \tag{32}$$

In this case, no matter how large the real RV is, there is no SDDA, i.e., $v_{r,\text{time}} = v_{r,\text{space}}$. The velocity obtained based on interferometric phase can be directly applied to the azimuth location of a moving target. However, if the real RV is outside the interval $[-V_T/2, V_T/2)$, it will suffer from TDDA. Therefore, the unambiguous velocity range in a Case I system is $[-V_T/2, V_T/2) = [-\lambda f_P/4, \lambda f_P/4)$. From (10) and (18), the real RV can be expressed as

$$\begin{cases} v_r = v_{r,\text{space}} + N_T V_T \\ d < 2v_a/f_P \end{cases} . \tag{33}$$

Since a ground moving target's RV is always limited, the TDDA integer $N_T$ won't be too large. Take an example from a real multichannel SAR system with typical parameters. The wavelength is $\lambda = 0.03\text{m}$, and the other parameters are the same as those in TABLE II in Section V. It can be calculated that $V_T = 12\text{m/s}$ and $V_S = 18\text{m/s}$. In order to show the ambiguity phenomenon, the estimated velocity is depicted in Fig. 5 based on the interferometric phase in image domain versus the real RV in a Case I system. It can be seen that the real ambiguity modulus in a Case I system is 12m/s, which is identical to the TDDA modulus. Fig. 2 illustrates the velocities after TDDA and SDDA of a moving target with $v_r = 17\text{m/s}$. It is indicated that, although there are two kinds of data samplings, the time sampling is ahead of the space sampling, so the target suffers from TDDA first. Due to $V_T < V_S$, $v_{r,\text{time}}$ after TDDA is always smaller than $V_S$, and will never suffer from SDDA.

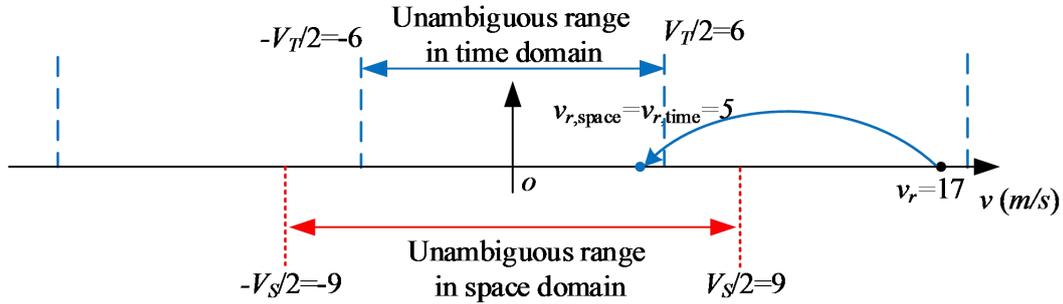

Fig. 2 Doppler ambiguity in a Case I system for a target with RV of 17m/s.

### 2) Case II

The condition of this case is

$$V_T \geq V_S \text{ and } V_T = kV_S, k = 1, 2, 3, \cdots. \tag{34}$$

From (9) and (16), this condition is equivalent to

$$d = 2kv_a / f_P, k = 1, 2, 3, \cdots. \tag{35}$$

The condition (35) is also the well-known DPCA condition [38], i.e., the spacing of the equivalent phase center is an integer multiple of the platform's flying distance between pulses, which leads to some convenience of target detection [39, 40]. Interestingly, the condition will also be beneficial for RV estimation herein. From (10) and (18), when $V_T \geq V_S$, if the moving target' RV is outside the interval $[-V_T/2, V_T/2)$, the TDDA arises as $v_r = v_{r,\text{time}} + N_T V_T$, while the existence of SDDA depends on whether $v_{r,\text{time}} \in \left[-V_S/2, V_S/2\right)$ or not. As we don't know which of them will arise, they can be represented as $v_{r,\text{time}} = v_{r,\text{space}} + N_S V_S$. For the former, $N_S = 0$, while for the latter, $N_S$ is a nonzero integer. Therefore, the real RV can be expressed as

$$\begin{cases} v_r = v_{r,\text{space}} + N_S V_S + N_T V_T \\ d > 2v_a / f_P \end{cases}. \tag{36}$$

Equation (36) is the so-called CTSDA, which is quite different from (33) in a Case I system. However, substituting (34) into (36) with the consideration of the special parameters for a Case II system, it will lead to

$$\begin{cases} v_r = v_{r,\text{space}} + N_{ST}V_S \\ d = 2kv_a/f_p, k=1,2,3,\cdots \end{cases}, \tag{37}$$

where $N_{ST} = N_S + kN_T$ is a new ambiguous integer. From (37) it is shown that the CTSDA problem in (36) is degenerated into an SDDA-only problem, and the unambiguous velocity range in a Case II system is $\left[-V_S/2, V_S/2\right) = \left[-\lambda v_a/(2d), \lambda v_a/(2d)\right)$. The estimated velocity is shown in Fig. 5 based on interferometric phase in image domain versus the real RV in a Case II system. The system parameters are the same as Case I, except for $d = 0.6\text{m}$. Then it can be calculated that $V_T = 12\text{m/s}$, $V_S = 6\text{m/s}$, and $d = 4v_a/f_P$, which satisfies the DPCA condition in (35) with $k = 2$. From Fig. 5 it can be seen that the real ambiguity modulus is 6m/s, which is identical to the SDDA modulus. Fig. 3 illustrates the velocities after TDDA and SDDA of a moving target with $v_r = 17\text{m/s}$. The target suffers from TDDA first, then the velocity changes to 5m/s with $N_T = 1$. Due to $V_S < V_T$, it will still suffer from SDDA, then the final estimated velocity is -1m/s with $N_S = 1$. Although there are two kinds of ambiguities, the real RV can be directly expressed as $v_r = v_{r,\text{space}} + 3V_S$ due to $V_T = 2V_S$.

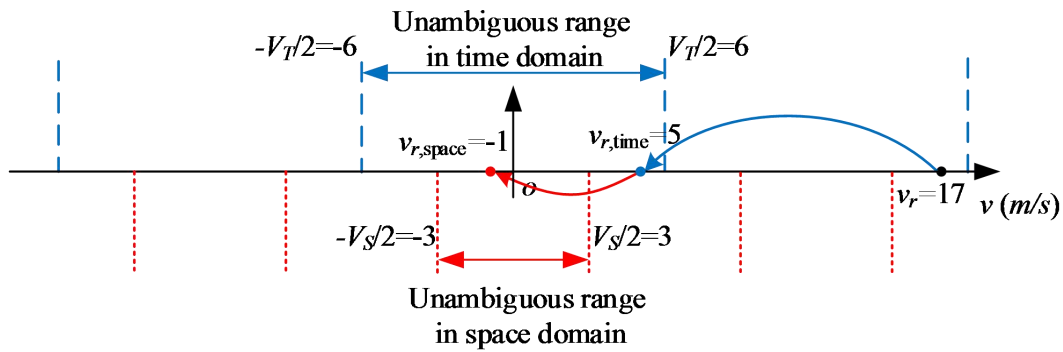

Fig. 3. Doppler ambiguity in a Case II system for a target with RV of 17m/s.

### 3) Case III

The condition of this case is

$$V_T > V_S \text{ and } V_T \neq kV_S, k=1,2,3,\cdots. \tag{38}$$

From (9) and (16), this condition is equivalent to

$$d > 2v_a/f_P \text{ and } d \neq 2kv_a/f_P, k=1,2,3,\cdots. \tag{39}$$

This is the common case in a multichannel SAR system. The unambiguous velocity range in a Case III system is $\left[-V_S/2, V_S/2\right) = \left[-\lambda v_a/(2d), \lambda v_a/(2d)\right)$. From (36) the CTSDA problem in Case III can be expressed as

$$\begin{cases} v_r = v_{r,\text{space}} + N_S V_S + N_T V_T \\ d > 2v_a/f_P, \, d \neq 2kv_a/f_P \end{cases}. \tag{40}$$

As SDDA is cascaded after TDDA and $V_T$ won't be too large in terms of the restricted PRF, the SDDA integer $N_S$ is always finite, i.e.,

$$N_S \in \left[\left[-V_T/2\right]_{V_S}, \left[V_T/2-1\right]_{V_S}\right], \tag{41}$$

where $\left[a\right]_b$ denotes the ambiguous integer of $a$ modulo $b$ as

$$\left[a\right]_b = \begin{cases} \left\lfloor \dfrac{a}{b} \right\rfloor, & \text{if } 0 \leq a - \left\lfloor \dfrac{a}{b} \right\rfloor b < \dfrac{b}{2} \\ \left\lfloor \dfrac{a}{b} \right\rfloor + 1, & \text{if } \dfrac{b}{2} \leq a - \left\lfloor \dfrac{a}{b} \right\rfloor b < b \end{cases}, \tag{42}$$

and here $\lfloor \cdot \rfloor$ is the floor operator. Take the SAR system in Case I for example, while $d$ changes to 0.4m to meet the condition in (39). It can be calculated that $V_T = 12$m/s, $V_S = 9$m/s, and $N_S \in \left[-1, 1\right]$. The estimated velocity is also depicted in Fig. 5 based on the interferometric phase in image domain versus the real RV in a Case III system. It can be observed that the variation tendency of the estimated velocity in this case is quite different from those in Case I and Case II, which are only folded by $V_T$ or $V_S$. The real ambiguity modulus of the estimated velocity in Case III is related to both $V_T$ and $V_S$. Fig. 4 illustrates the velocities after TDDA and SDDA of a moving target with $v_r = 17$m/s. The velocity after TDDA is 5m/s, which is still larger than $V_S/2$. Hence, it will suffer from SDDA, and the final estimated velocity changes to -4m/s.

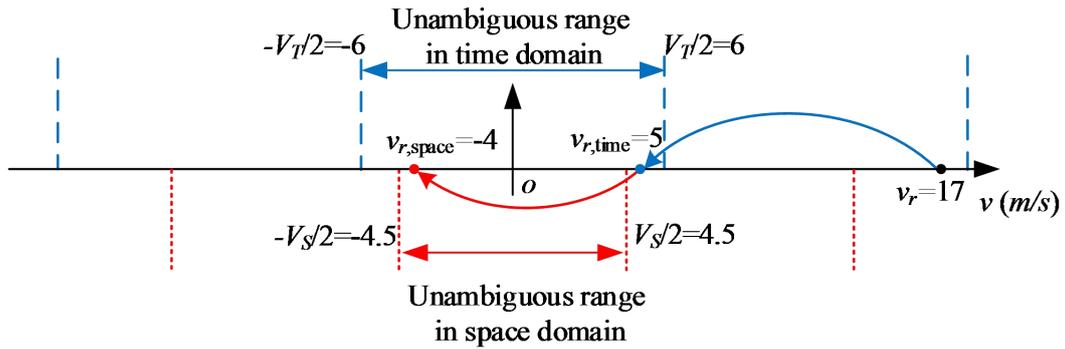

Fig. 4 Doppler ambiguity in a Case III system for a target with RV of 17m/s.

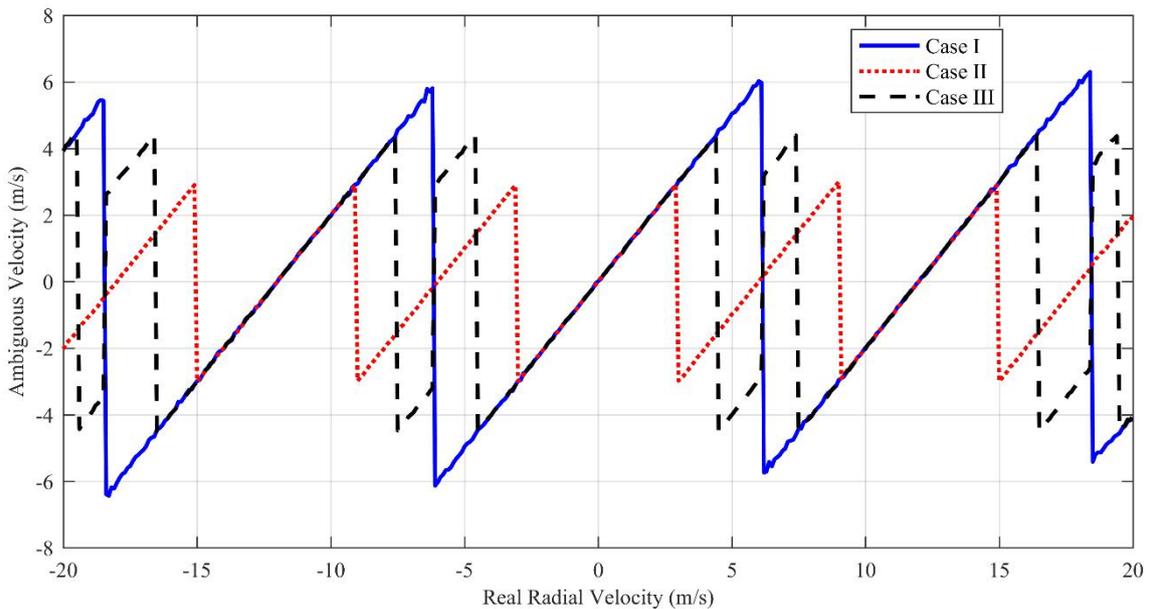

Fig. 5 Estimated velocity based on interferometric phase in image domain.

In line with the above analysis, the Doppler ambiguity in a multichannel SAR system can be summarized in TABLE I in accordance with the proposed system classification. Obviously, a moving target may have different ambiguities in different systems, which are determined by its motion parameters as well as the system parameters. The unambiguous velocity ranges in the three cases are related to the system parameters PRF $f_P$, channel spacing $d$ and platform velocity $v_a$. For convenience, define the determinable velocity size as the interval length of the unambiguous velocity range. Based on (32), (35) and (39), the influences of the system parameters on the determinable velocity size are described in Fig. 6. When $d$ and $v_a$ are invariant, the determinable velocity size versus $f_P$ is shown in Fig. 6(a). It can be observed that, when $f_P$ is smaller than $2v_a/d$, the system belongs to Case I, and the determinable velocity size is increasing with the PRF monotonically. When $f_P$ surpasses $2v_a/d$, the determinable velocity size remains $\lambda v_a/d$, because the system changes to Case II or III. Similarly, from Fig. 6(b) and Fig. 6(c), it is shown that with the increase of $d$ or $v_a$, the system will change from Case II or III to Case I, and the maximum determinable velocity size is $\lambda f_P/2$. In conclusion, increasing the PRF, platform velocity or decreasing the channel spacing separately can only improve the unambiguous velocity range to a limited extent.

TABLE I System Cases and Time-Space Doppler Ambiguity.

| System cases | Case I | Case II | Case III |
|---|---|---|---|
| Conditions | $d < 2v_a/f_P$ | $d = 2kv_a/f_P, k=1,2,3,\cdots$ | $d > 2v_a/f_P$ and $d \neq 2kv_a/f_P$ |
| Unambiguous velocity range | $\left[-\lambda f_P/4, \lambda f_P/4\right)$ | $\left[-\lambda v_a/(2d), \lambda v_a/(2d)\right)$ | $\left[-\lambda v_a/(2d), \lambda v_a/(2d)\right)$ |
| TDDA | √ | √ | √ |
| SDDA | × | √ | √ |
| Velocity after TDDA | $v_{r,\text{space}}$ | $v_{r,\text{space}} + N_S V_S$ | $v_{r,\text{space}} + N_S V_S$ |
| Real RV | $v_{r,\text{space}} + N_T V_T$ | $v_{r,\text{space}} + \left(N_S + kN_T\right)V_S$ | $v_{r,\text{space}} + N_S V_S + N_T V_T$ |
| Azimuth shift | $-R_0 v_{r,\text{space}}/v_a$ | $-R_0\left(v_{r,\text{space}} + N_S V_S\right)/v_a$ | $-R_0\left(v_{r,\text{space}} + N_S V_S\right)/v_a$ |
| $\Delta x_{\max}$ | $\lambda f_P R_0/4v_a$ | $\lambda f_P R_0/4v_a$ | $\lambda f_P R_0/4v_a$ |

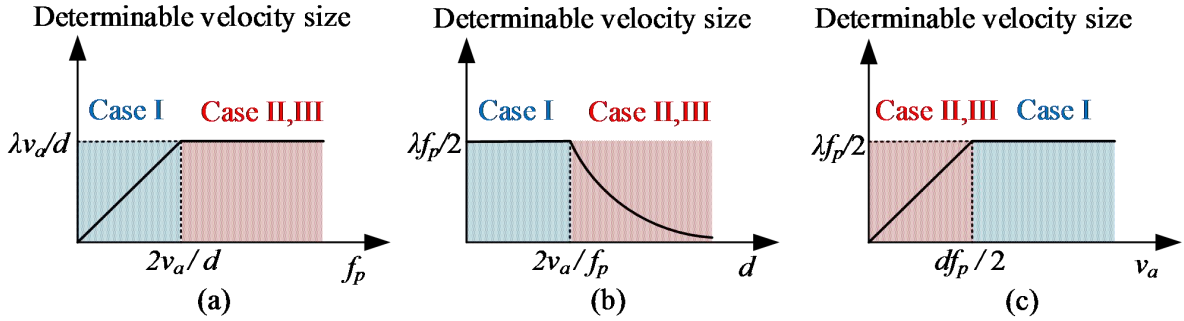

Fig. 6 The relationship between determinable velocity size and system parameters.

## IV. Multi-Frequency SAR for RV Retrieval

From (16) it is shown that the SDDA velocity modulus $V_S$ is related to the channel spacing and platform velocity, thus the SDDA problem in (18) can be solved by NULA-SAR [23] and DS-SAR [24] based on CRT. However, they can only obtain the ambiguous RV $v_{r,\text{time}}$ after TDDA rather than the real RV $v_r$. From (9) and (16), it is shown that the TDDA velocity modulus $V_T$ and the SDDA velocity modulus $V_S$ are both in correlation with wavelength $\lambda$. Therefore, the MF-SAR [32] can be used for the real RV retrieval. In [32], only SDDA as well as "azimuth location ambiguity" is discussed based on the conventional CRT. In this section, an MF-SAR is proposed to obtain the real RV by solving TDDA and SDDA jointly. Assume the radar transmits signals with $L$ different carrier wavelengths $\lambda_i, i = 1, 2, \cdots, L$. The $L$ estimated velocities $v_{r,\text{space},i}, i = 1, 2, \cdots, L$ can be obtained based on interferometric phase from all the wavelengths.

*A. MF-SAR for a Case I System*

From (33) the ambiguity problem in a Case I system with multi-frequencies can be expressed as

$$\begin{cases} v_r = v_{r,\text{space},i} + N_{T,i} V_{T,i}, i = 1, 2, \cdots, L \\ V_{T,i} < V_{S,i} \end{cases},$$ (43)

which is the conventional CRT problem and can be solved by the closed-form robust CRT [33, 34]. The determinable velocity size of MF-SAR is

$$\begin{aligned} v_{s,1} &= \text{lcm}\left(V_{T,1}, V_{T,2}, \cdots, V_{T,L}\right) \\ &= \text{lcm}\left(\frac{\lambda_1 f_P}{2}, \frac{\lambda_2 f_P}{2}, \cdots, \frac{\lambda_L f_P}{2}\right), \end{aligned}$$ (44)

where $\text{lcm}(\cdot)$ stands for the least common multiple. Thus, the maximum determinable velocity of a Case I system is

$$\begin{aligned} \left|v_{\max,1}\right| &= \frac{1}{2} \text{lcm}\left(V_{T,1}, V_{T,2}, \cdots, V_{T,L}\right) \\ &= \frac{1}{2} \text{lcm}\left(\frac{\lambda_1 f_P}{2}, \frac{\lambda_2 f_P}{2}, \cdots, \frac{\lambda_L f_P}{2}\right), \end{aligned}$$ (45)

where 1/2 is multiplied because there are two possible different motion directions of $v_r$, i.e., toward and backward to the radar.

*B. MF-SAR for a Case II System*

From (37) the ambiguity problem in a Case II system with multi-frequencies can be expressed as

$$\begin{cases} v_r = v_{r,\text{space},i} + N_{ST,i} V_{S,i}, i = 1, 2, \cdots, L \\ N_{ST,i} = N_{S,i} + k N_{T,i} \\ V_{T,i} = k V_{S,i}, k = 1, 2, 3 \cdots \end{cases},$$ (46)

which is the conventional CRT problem for $N_{ST,i}$ instead of $N_{T,i}$ and $N_{S,i}$, and can be solved by the closed-form robust CRT [33, 34]. The determinable velocity size of MF-SAR is

$$v_{s,\mathrm{II}} = \mathrm{lcm}\left(V_{S,1}, V_{S,2}, \cdots, V_{S,L}\right)$$
$$= \mathrm{lcm}\left(\frac{\lambda_1 v_a}{d}, \frac{\lambda_2 v_a}{d}, \cdots, \frac{\lambda_L v_a}{d}\right), \tag{47}$$

and thus, the maximum determinable velocity is

$$\left|v_{\max,\mathrm{II}}\right| = \frac{1}{2}\mathrm{lcm}\left(V_{S,1}, V_{S,2}, \cdots, V_{S,L}\right)$$
$$= \frac{1}{2}\mathrm{lcm}\left(\frac{\lambda_1 v_a}{d}, \frac{\lambda_2 v_a}{d}, \cdots, \frac{\lambda_L v_a}{d}\right). \tag{48}$$

After estimating the real RV, the velocity after TDDA can be calculated by (10), which can be used for relocation.

### C.  MF-SAR for a Case III System

For  $i = 1, 2, \cdots, L$ , from (40) it is able to derive  $v_r = v_{r,\mathrm{space},i} + N_{S,i}V_{S,i} + N_{T,i}V_{T,i}$ . This is quite different from the conventional CRT problem in Case I and Case II systems. Considering the relationship between $V_{T,i}$ and $V_{S,i}$, the CTSDA problem in Case III can be expressed as

$$\begin{cases} v_r = v_{r,\mathrm{space},i} + N_{S,i}V_{S,i} + N_{T,i}V_{T,i}, & i = 1, 2, \cdots, L \\ V_{T,i} > V_{S,i} \text{ and } V_{T,i} \neq k V_{S,i} \\ \dfrac{V_{T,i}}{V_{S,i}} = \dfrac{df_P}{2v_a} = \dfrac{p}{q} \\ -V_{T,i}/2 \leq v_{r,\mathrm{space},i} + N_{S,i}V_{S,i} < V_{T,i}/2 \end{cases}, \tag{49}$$

where the third equation is from the definitions of  $V_T$  and  $V_S$ , and $p$ and $q$ are co-prime. From (49), one can see that two levels of modulo operations are involved, where the RV is first taken a modulo with a blind velocity, $V_{T,i}$, in time domain and then its remainder is taken another modulo with a blind velocity, $V_{S,i}$, in space domain, and $V_{S,i} < V_{T,i}$. This effect is called a double remaindering problem in this paper. For this double remaindering problem, the first question is what the unambiguous velocity range is.

**Theorem 1:** If  $v_r$  is fallen in the range of  $\left[-v_{\mathrm{lb}}/2, v_{\mathrm{lb}}/2\right)$ , where  $v_{\mathrm{lb}} = \mathrm{lcm}\left(V_{S,1}, V_{S,2}, \cdots, V_{S,L}\right)/q$ , it can be uniquely determined in the double remaindering problem.

This theorem is proved in Appendix A, and provides a lower bound  $v_{\mathrm{lb}}$  for the determinable velocity size. One can see from the proof of Theorem 1 that when the sufficient condition in Theorem 1 is satisfied, the double remaindering problem is degenerated to the conventional CRT problem with moduli $V_{S,i}/q$, $i = 1, 2, \cdots, L$, and remainders $\zeta_i, i = 1, 2, \cdots, L$ as defined in (58), and thereby it can be solved by the closed-form robust CRT [33, 34].

Unfortunately, the condition in Theorem 1 may not be necessary for the uniqueness of the solution, and the unambiguous velocity range can even reach $\left[ -v_{s,\mathrm{III}}/2, v_{s,\mathrm{III}}/2 \right)$ for some system parameters, where $v_{s,\mathrm{III}} = \mathrm{lcm}\left( V_{T,1}, V_{T,2}, \cdots, V_{T,L} \right)$, as we shall see in the following examples. Here we give an example to analyze the determinable velocity size in MF-SAR for a Case III system. Without loss of generality, two wavelengths $\lambda_1$ and $\lambda_2$ are used. $\left( \lambda_1, \lambda_2 \right)$ take the values varying from (0.02m,0.03m) to (0.11m,0.12m) with a step length 0.01m, and the other system parameters are the same as those in TABLE II. Then, the TDDA and SDDA moduli with $\lambda_1$ and $\lambda_2$ can be calculated by (9) and (16), respectively, which are listed in Appendix B. For each of wavelength pairs $\left( \lambda_1, \lambda_2 \right)$, let the target's RV changes as $\{0, -1, 1, -2, 2, \cdots\}$ m/s one by one, then the pair of the residues can be calculated from (49). Once the pair of the residues are the same as the one of the former pairs, the corresponding RV of this time is the maximum determinable velocity. Then, the determinable velocity size can be calculated as twice the maximum determinable velocity. As shown in Fig. 7, the lcm($V_{T,1}$, $V_{T,2}$) and lcm($V_{S,1}$, $V_{S,2}$) are monotonically increasing, while the determinable velocity size in MF-SAR changes irregularly for a Case III system. Sometimes it reaches the upper bound lcm($V_{T,1}$, $V_{T,2}$), sometimes it drops to the lower bound, and sometimes it is in the middle of the upper and the lower bounds. Some of the detailed examples of the determinable velocity size can be found in Appendix B. For example, when $\lambda_1 = 0.05\mathrm{m}$, $\lambda_2 = 0.06\mathrm{m}$, i.e., $V_{T,1}$=20m/s, $V_{S,1}$=15m/s, $V_{T,2}$=24m/s, $V_{S,2}$=18m/s, the determinable velocity size is $v_{s,\mathrm{III}} = 120\mathrm{m/s}$, which satisfies the common possible ground moving target's RV range and will be used in Section V for numerical experiments.

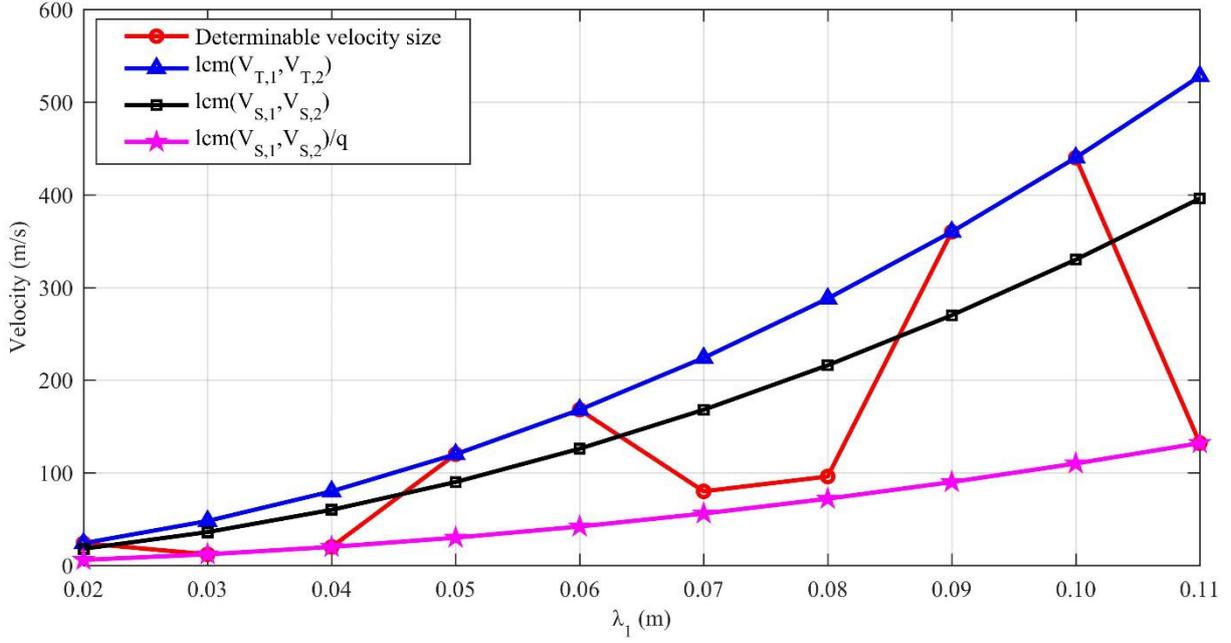

Fig. 7 Determinable velocity size in MF-SAR for a Case III system.

Note that, when the sufficient condition in Theorem 1 is not satisfied, the closed-form robust CRT [33, 34] may not be suitable for the double remaindering problem. In this case, we propose another searching based RV reconstruction method based on the existing robust CRT [35, 36] for the CTSDA problem in a Case III system. Assume $\lambda_1 < \lambda_2 < \cdots < \lambda_L$, then $V_{T,1} < V_{T,2} < \cdots < V_{T,L}$ and $V_{S,1} < V_{S,2} < \cdots < V_{S,L}$. Suppose the estimated velocity $v'_{r,\text{space},i}$ based on interferometric phase has an uncertain error $\varepsilon_i$, i.e., $v_{r,\text{space},i} = v'_{r,\text{space},i} + \varepsilon_i$, and $|\varepsilon_i| \le \xi_e$, where $\xi_e$ is the error bound. Then the real RV can be expressed as

$$v_r = \varepsilon_i + v'_{r,\text{space},i} + N_{S,i}V_{S,i} + N_{T,i}V_{T,i}, i = 1, 2, \cdots, L \ . \tag{50}$$

For each $i$ with $2 \le i \le L$, let us define

$$S_i = \left\{ \begin{array}{l} \left( \bar{N}_{T,1}, \bar{N}_{S,1}, \bar{N}_{T,i}, \bar{N}_{S,i} \right) = \\ \arg \min_{\substack{ \hat{N}_{T,1} \in \left[ \left[ -v_{s,\text{III}}/2 \right]_{V_{T,1}}, \left[ v_{s,\text{III}}/2 - 1 \right]_{V_{T,1}} \right] \\ \hat{N}_{S,1} \in \left[ \left[ -V_{T,1}/2 \right]_{V_{S,1}}, \left[ V_{T,1}/2 - 1 \right]_{V_{S,1}} \right] \\ \hat{N}_{T,i} \in \left[ \left[ -v_{s,\text{III}}/2 \right]_{V_{T,i}}, \left[ v_{s,\text{III}}/2 - 1 \right]_{V_{T,i}} \right] \\ \hat{N}_{S,i} \in \left[ \left[ -V_{T,i}/2 \right]_{V_{S,i}}, \left[ V_{T,i}/2 - 1 \right]_{V_{S,i}} \right] }} \left| v'_{r,\text{space},i} + \hat{N}_{S,i}V_{S,i} + \hat{N}_{T,i}V_{T,i} - \left( v'_{r,\text{space},1} + \hat{N}_{S,1}V_{S,1} + \hat{N}_{T,1}V_{T,1} \right) \right| \end{array} \right\}, \tag{51}$$

$$\text{s.t.} \begin{cases} -V_{T,i}/2 - \xi_e \le v'_{r,\text{space},i} + \hat{N}_{S,i}V_{S,i} < V_{T,i}/2 + \xi_e \\ -V_{T,1}/2 - \xi_e \le v'_{r,\text{space},1} + \hat{N}_{S,1}V_{S,1} < V_{T,1}/2 + \xi_e \end{cases}$$

where $v_{s,\text{III}}$ is the determinable velocity size in MF-SAR for a Case III system, and can be determined like the example above for a real system. Recall $[a]_b$ is defined in (42).

Furthermore, let us define $S_{i,T,1}$ and $S_{i,S,1}$ as the sets of all the first components $\overline{N}_{T,1}$ and all the second components $\overline{N}_{S,1}$ of the pairs $\left( \overline{N}_{T,1}, \overline{N}_{S,1}, \overline{N}_{T,i}, \overline{N}_{S,i} \right)$ in $S_i$, respectively, i.e.,

$$\begin{cases} S_{i,T,1} = \left\{ \overline{N}_{T,1} \middle| \left( \overline{N}_{T,1}, \overline{N}_{S,1}, \overline{N}_{T,i}, \overline{N}_{S,i} \right) \in S_i \right\}, i=2,\cdots,L \\ S_{i,S,1} = \left\{ \overline{N}_{S,1} \middle| \left( \overline{N}_{T,1}, \overline{N}_{S,1}, \overline{N}_{T,i}, \overline{N}_{S,i} \right) \in S_i \right\}, i=2,\cdots,L \end{cases}, \tag{52}$$

and define

$$\begin{cases} S_T = \bigcap_{i=2}^{L} S_{i,T,1} \\ S_S = \bigcap_{i=2}^{L} S_{i,S,1} \end{cases}. \tag{53}$$

Then, if each of the sets $S_T$ and $S_S$ contains one and only one element $N_{T,1}$ and $N_{S,1}$, respectively, i.e., $S_T = \left\{ N_{T,1} \right\}$ and $S_S = \left\{ N_{S,1} \right\}$, and furthermore, if $\left( N_{T,1}, N_{S,1}, \overline{N}_{T,i}, \overline{N}_{S,i} \right) \in S_i$, then $\overline{N}_{T,i} = N_{T,i}$, $\overline{N}_{S,i} = N_{S,i}$ for $2 \leq i \leq L$, where $N_{T,i}$ and $N_{S,i}$, $1 \leq i \leq L$, are the solutions of (49). When the ambiguity integers are solved, the real RV can be estimated as

$$\hat{v}_r = \frac{1}{L} \sum_{i=1}^{L} \left( v'_{r,\text{space},i} + N_{S,i} V_{S,i} + N_{T,i} V_{T,i} \right). \tag{54}$$

It is worth mentioning that, although this searching based method has a 4-D search of $\hat{N}_{T,i}$, $\hat{N}_{S,i}$, $\hat{N}_{T,1}$ and $\hat{N}_{S,1}$, it doesn't have too high computational complexity because only a limited number of possible ambiguity integers are needed to be searched as what is discussed in Section V.

## V. NUMERICAL EXPERIMENTS AND PERFORMANCE ANALYSIS

### A. Numerical Results of the Proposed MF-SAR

Since the Doppler ambiguity problem in Cases I and II with multi-frequencies is the conventional CRT problem, the simulation analyses of them are not discussed in this section for simplicity. To demonstrate the effectiveness of MF-SAR for a Case III system, some numerical simulations are presented in this section. The parameters of a Case III system are listed in TABLE II. It can be calculated that $V_{T,1}$=20m/s, $V_{S,1}$=15m/s, $V_{T,2}$=24m/s, $V_{S,2}$=18m/s, and the unambiguous velocity range in this MF-SAR is $[-60,60)$m/s . From (42) and (51), the ambiguous integers are $N_{S,i} \in \{-1,0,1\}, i=1,2$ , $N_{T,1} \in \{-3,-2,-1,0,1,2,3\}$ and $N_{T,2} \in \{-2,-1,0,1,2\}$ , thus the 4-D search in the searching based method doesn't have too high computational complexity in the application of GMTI. Five moving targets, T1, T2, T3, T4 and T5, produced by simulation are added into the real measured SAR raw data of the static scene with radial velocities 8.36m/s, 13.46m/s, 17.01m/s, -11.03m/s, and -16.87m/s on the roads, respectively. The clutter-to-noise ratio in the SAR image is set as 20 dB, which means the existence of interferometric phase noise before performing VSAR. The five moving targets all suffer from Doppler ambiguities when $\lambda_1$ or $\lambda_2$ is used alone, while they are all in the unambiguous range of MF-SAR.

The clutter is suppressed and the targets are detected via VSAR. Fig. 8(a) and Fig. 8(b) show the static scene SAR image marked with the detected targets with wavelengths $\lambda_1$ and $\lambda_2$, respectively. The signal-to-clutter-noise ratios of the five moving targets after clutter suppression are about 14.5dB, 11.3dB, 8.7dB, 7.9dB and 10.4dB, respectively, in wavelength $\lambda_1$ , while 14.7dB, 11.1dB, 8.4dB, 14.1dB and 9.6dB, respectively, in wavelength $\lambda_2$ . It indicates the shifts of the moving targets in the azimuth and all five moving targets are imaged out of the roads. However, since the TDDA moduli in wavelengths $\lambda_1$ and $\lambda_2$ are different, the target's velocities after TDDA will be different, too. So it can be seen that the azimuth shifts of targets are varied in Fig. 8(a) and Fig. 8(b). Furthermore, targets T1, T3 and T4 are zoomed-in with yellow rectangles. Interestingly, T1 and T3 belong to Type I and Type II targets, respectively, for both wavelengths $\lambda_1$ and $\lambda_2$ . However, T4 belongs to Type II for wavelength $\lambda_1$ , but belongs to Type I for wavelength $\lambda_2$ , due to its different TDDA integers in the two wavelengths. The radial velocities estimated by VSAR are listed in TABLE III, which are all ambiguous. When $M$=8, the velocity resolutions are about 2.14m/s and 2.57m/s for wavelengths $\lambda_1$ and $\lambda_2$, respectively. For simplicity, only one point target is contained in a pixel in our simulation, so the RV estimation accuracy via the FFT-based VSAR can be improved by zero-padding FFT. In our experiments, 1000 times of interpolation, i.e., the 8000-point zero-padding FFT, is implemented in TABLE III to estimate the ambiguous RV suffered by TDDA and/or SDDA. Note that, in the practical applications, if two or more targets with RV difference smaller than the velocity resolution in a pixel, they cannot be resolved and the RV estimated results will not have the high accuracy as shown in TABLE III. Then the searching based method can be used to retrieve the real RV by solving the double remaindering problem. From TABLE III, it can be seen that the five targets have different ambiguous integers, and all of the radial velocities can be estimated accurately by the searching based method. Furthermore, in order to demonstrate Theorem 1, the closed-form robust CRT method in [33] is used to retrieve the radial velocities with moduli $V_{S,1}/q = 5$ and $V_{S,2}/q = 6$. The estimated results are listed in TABLE III. According to Theorem 1, in this MF-SAR, the velocities in the range of $[-15,15)$ m/s can be uniquely determined by the closed-form robust CRT. From TABLE III, it is shown that the closed-form robust CRT can retrieve the real radial velocities accurately for T1, T2 and T4, whose velocities are all in the range of $[-15,15)$ m/s, while the estimated results of T3 and T5 are erroneous since they are not in the range demonstrated by Theorem 1. After estimating the radial velocities by the searching based method, the velocities after TDDA can be calculated to obtain the azimuth shifts based on (28). The azimuth shifts of targets T1, T2, T3, T4, T5 are -697.4250m, 545.8m, 248.7833m, -753.4583m, -260.8333m in wavelength $\lambda_1$, and -697.4250m, 879.1333m, 582.1167m, 913.2083m, -594.1667m in wavelength $\lambda_2$, respectively. The relocation results of the five moving targets are illustrated in Fig. 9 and Fig. 10, from which we can see that the five targets are all accurately located on the roads.

TABLE II System parameters in MF-SAR for a Case III system.

| Parameters | Value |
| --- | --- |
| Channel spacing ($d$) | 0.4 m |
| Wavelength ($\lambda$) | $\{0.05\text{m}, 0.06\text{m}\}$ |
| Platform velocity ($v_a$) | 120 m/s |
| PRF ($f_P$) | 800 Hz |
| Bandwidth ($B$) | 80 MHz |
| Sampling frequency ($f_s$) | 100 MHz |
| Center range ($R_0$) | 10000 m |
| Pulse duration ($T$) | 2.25μs |

| Antenna number ($M$) | 8 |
|---|---|

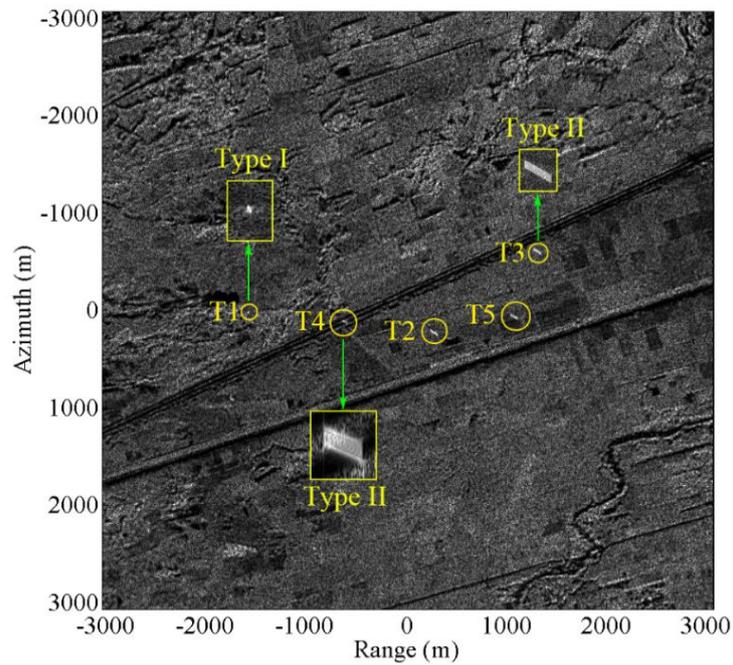

(a) $\lambda_1$

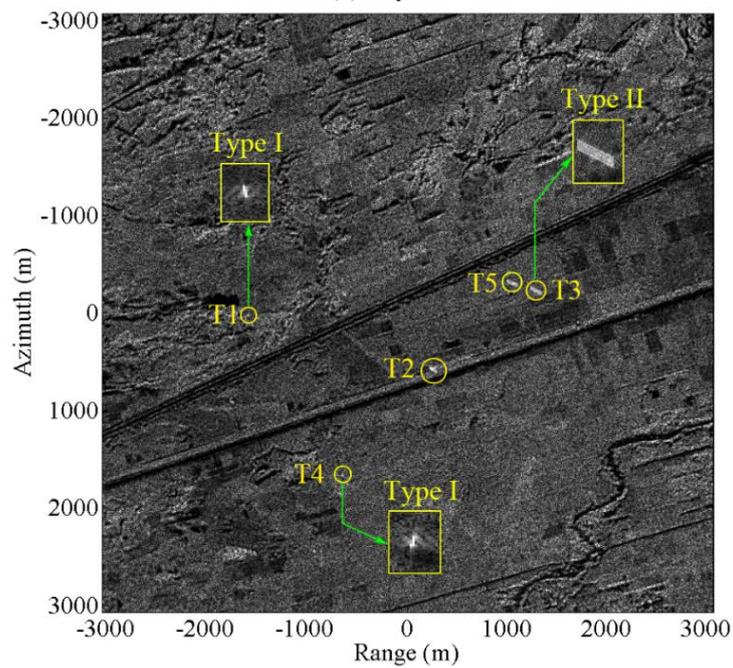

(b) $\lambda_2$

Fig. 8 Scene imaging results marked with the detected targets.

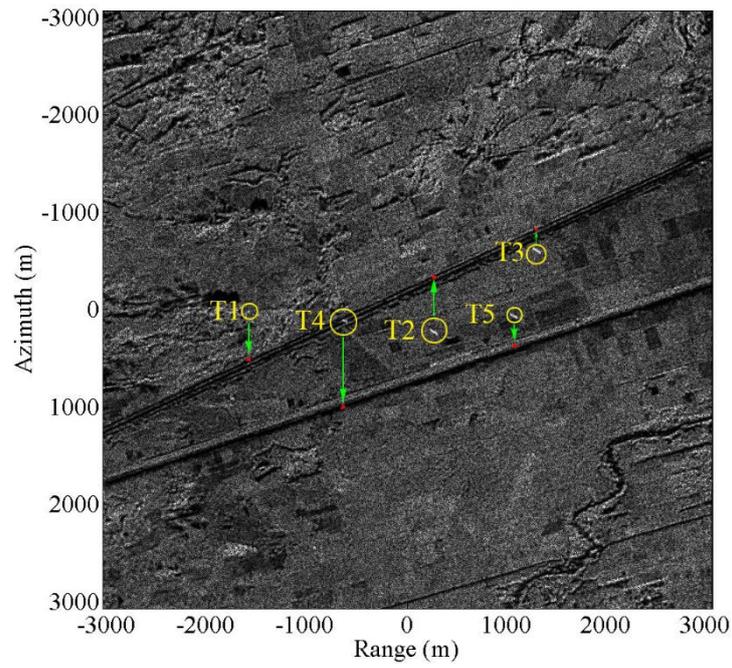

Fig. 9 Relocation of moving targets in a Case III system with $\lambda_1$.

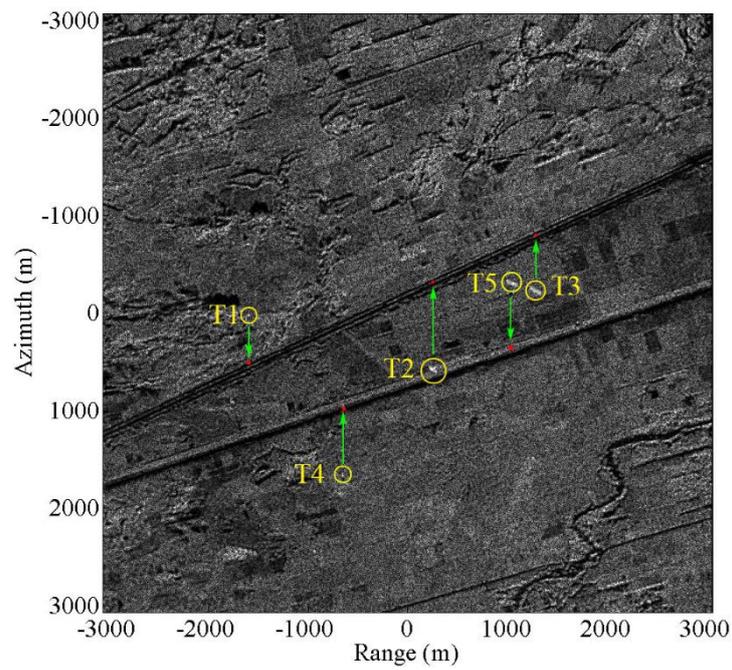

Fig. 10 Relocation of moving targets in a Case III system with $\lambda_2$.

TABLE III Moving target's RV and estimated results.

| Target number | Real RV | Estimated results by VSAR | | | Searching based method | | | | | | Closed-form robust CRT |
|---|---|---|---|---|---|---|---|---|---|---|---|
| | $v_r$ (m/s) | $v_{r,\text{space},1}$ (m/s) | $v_{r,\text{space},2}$ (m/s) | $N_{T,1}$ | $N_{S,1}$ | $N_{T,2}$ | $N_{S,2}$ | $\hat{v}_r$ (m/s) | | | $\hat{v}_r$ (m/s) |
| T1 | 8.36 | -6.5791 | 8.3173 | 0 | 1 | 0 | 0 | 8.3691 | | | 8.3691 |
| T2 | 13.46 | -6.4708 | 7.3716 | 1 | 0 | 1 | -1 | 13.4504 | | | 13.4504 |
| T3 | 17.01 | -3.1730 | -6.7979 | 1 | 0 | 1 | 0 | 17.0146 | | | -12.9855 |
| T4 | -11.03 | -5.8834 | 6.9664 | -1 | 1 | 0 | -1 | -10.9585 | | | -10.9585 |
| T5 | -16.87 | 3.1043 | 7.1790 | -1 | 0 | -1 | 0 | -16.8584 | | | 13.1417 |

## B. Performance Analysis for the Proposed Searching Based Reconstruction Method

In this section, some simulation results are provided to illustrate the searching based reconstruction method performance. The system parameters in MF-SAR for Case III are the same as those in TABLE II. The unknown RV of a target is chosen uniformly at random from the unambiguous range $[-60, 60]$ m/s, and the ambiguous velocities can be calculated by (49) accurately. To describe the estimation error in the real application, the uniformly distributed errors between $[-\xi_e, \xi_e]$ are added on the accurate ambiguous velocity. The root mean square error (RMSE) of the estimated velocities by the searching based reconstruction method versus $\xi_e$ is shown in Fig. 11, where $\xi_e$ changes from 1m/s to 0m/s with a step length -0.05m/s, and 10000 Monte Carlo trials are implemented for each of them. The RMSE is calculated by $E_{\text{RMSE}} = \sqrt{\sum_i^K (\hat{v}_{ri} - v_r)^2 \Big/ K}$, where $K$ is the number of the Monte Carlo trials. It shows that, as $\xi_e$ decreases, the RMSE of the estimated velocities decreases rapidly. When the error bound is lower than 0.5m/s, the RMSE becomes smaller than 0.2m/s. From the simulation results above, it can be concluded that the proposed searching based reconstruction method can retrieve the real RV robustly when the ambiguous velocities are estimated based on the interferometric phase in a certain error range. As a remark, the threshold $\xi_e = 0.5$ is a sufficient condition to the robust estimation of the RV by using the robust CRT. When this sufficient condition is not satisfied, the CRT based reconstruction may not be robust anymore, i.e., the reconstruction error may immediately become large as illustrated in Fig. 11. In addition, the sufficient and necessary condition of the double remaindering problem is still an open problem and deserved to be studied in the future.

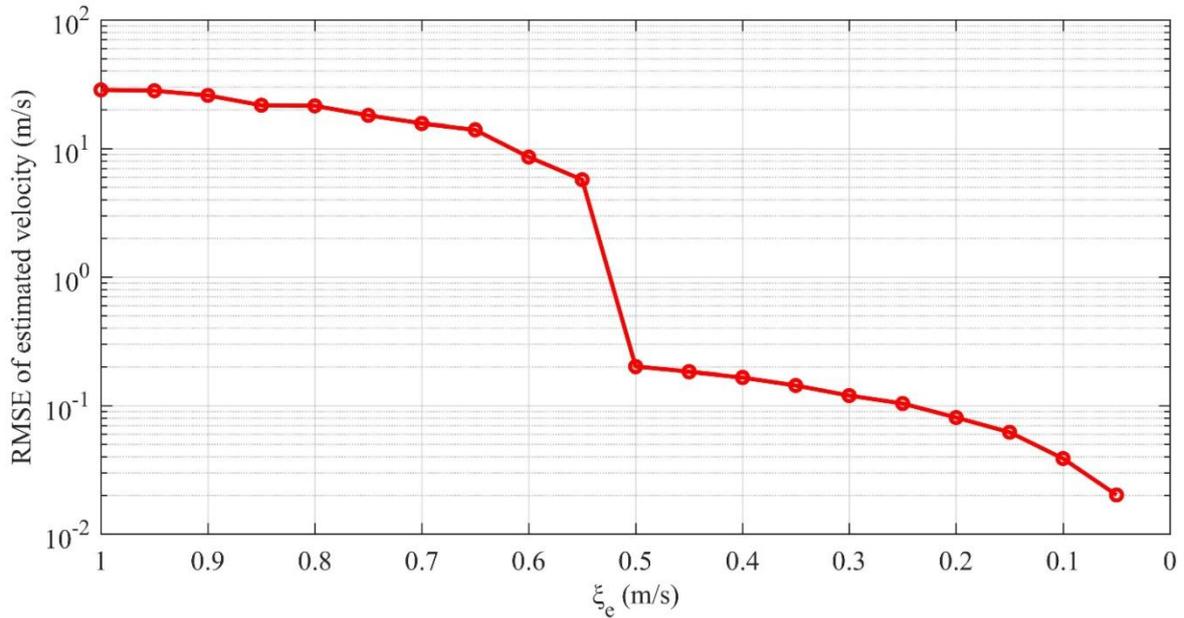

Fig. 11 RMSE of the estimated velocity by the searching based reconstruction method.

## VI. CONCLUSIONS

In this paper, TDDA and SDDA are derived for multichannel SAR moving targets in range-compressed domain, RD domain and image domain, respectively. In accordance with the analyses of the relationship between TDDA and SDDA, it is indicated that SAR systems can be divided into three cases. For Case I, only TDDA will arise. For Cases II and III, the CTSDA arises, that is, TDDA in each channel arises first and then SDDA among multi-channels arises subsequently. Then, an MF-SAR is proposed for RV estimation by solving the ambiguity problem based on CRT. For Cases I and II, the RV can be uniquely retrieved by the closed-form robust CRT. In Case III, the CTSDA problem is different from the conventional CRT problem with multi-frequencies, which is called a double remaindering problem in this paper. A sufficient condition is derived for the uniqueness of a solution, under which the double remaindering problem can be solved by the closed-form robust CRT. When the sufficient condition is not satisfied, a searching based reconstruction method is proposed. Based on our numerical experiments and performance analysis, it is validated that the proposed MF-SAR can well accomplish the CTSDA de-ambiguity and obtain the unambiguous RV via the multiple ambiguous radial velocities in space domain with respect to multiple frequencies.

From the third equation in (49), it is possible to obtain

$$pV_{S,i} = qV_{T,i}. \tag{55}$$

Substitute (55) into the first equation in (49) and then

$$v_r = v_{r,\text{space},i} + \left(qN_{S,i} + pN_{T,i}\right)\frac{V_{S,i}}{q}, \quad i = 1, 2, \cdots, L. \tag{56}$$

That is,

$$v_r = v_{r,\text{space},i} \bmod \left(\frac{V_{S,i}}{q}\right), \quad i = 1, 2, \cdots, L. \tag{57}$$

Let

$$\zeta_i = \left\langle v_{r,\text{space},i} \right\rangle_{\frac{V_{S,i}}{q}}, \tag{58}$$

where $\left\langle a \right\rangle_b$ denotes the absolutely least remainder [41] of $a$ modulo $b$ in this paper. In other words, for real numbers $a$ and $b \geq 1$,

$$\left\langle a \right\rangle_b = \begin{cases} a - \left\lfloor \dfrac{a}{b} \right\rfloor b, & \text{if } 0 \leq a - \left\lfloor \dfrac{a}{b} \right\rfloor b < \dfrac{b}{2} \\ a - \left\lfloor \dfrac{a}{b} \right\rfloor b - b, & \text{if } \dfrac{b}{2} \leq a - \left\lfloor \dfrac{a}{b} \right\rfloor b < b \end{cases}. \tag{59}$$

Then,

$$v_r = \zeta_i \bmod \left(\frac{V_{S,i}}{q}\right), \quad i = 1, 2, \cdots, L. \tag{60}$$

Therefore, in the MF-SAR system, $v_r$ can be uniquely determined in the range of $v_r \in \left[-v_{\text{lb}}/2, v_{\text{lb}}/2\right)$ [42], where $v_{\text{lb}} = \text{lcm}\left(V_{S,1}/q, V_{S,2}/q, \cdots, V_{S,L}/q\right) = \text{lcm}\left(V_{S,1}, V_{S,2}, \cdots, V_{S,L}\right)/q$. In particular, when $q = 1$, (55) will be converted into $pV_{S,i} = V_{T,i}$, and (49) coincides with a Case II system with a determinable velocity size $\text{lcm}\left(V_{S,1}, V_{S,2}, \cdots, V_{S,L}\right)$.

TABLE IV. Determinable velocity size in Case III systems with different system parameters.

| $(\lambda_1,\lambda_2)$ (m) | $(V_{T,1},V_{S,1})$ (m/s) | $(V_{T,2},V_{S,2})$ (m/s) | $\mathrm{lcm}(V_{S,1},V_{S,2})/q$ (m/s) | Determinable velocity size (m/s) | $\mathrm{lcm}(V_{T,1},V_{T,2})$ (m/s) |
|---|---|---|---|---|---|
| (0.02,0.03) | (8,6) | (12,9) | 6 | 24 | 24 |
| (0.03,0.04) | (12,9) | (16,12) | 12 | 12 | 48 |
| (0.04,0.05) | (16,12) | (20,15) | 20 | 20 | 80 |
| (0.05,0.06) | (20,15) | (24,18) | 30 | 120 | 120 |
| (0.06,0.07) | (24,18) | (28,21) | 42 | 168 | 168 |
| (0.07,0.08) | (28,21) | (32,24) | 56 | 80 | 224 |
| (0.08,0.09) | (32,24) | (36,27) | 72 | 96 | 288 |
| (0.09,0.10) | (36,27) | (40,30) | 90 | 360 | 360 |
| (0.10,0.11) | (40,30) | (44,33) | 110 | 440 | 440 |
| (0.11,0.12) | (44,33) | (48,36) | 132 | 132 | 528 |

ACKNOWLEDGMENT

The first three authors would like to thank Dr. Weiran Huang in Tsinghua University for his constructive comments and suggestions to improve the quality of this paper. This work was supported in part by the National Natural Science Foundation of China under Grant 61671061.